\documentclass[mathleft
]{an}
\usepackage{graphicx}
\usepackage{times}
\begin{document}

% The following seven commands are intended for editorial usage and should be ignored by
% the author(s).
%\Pagespan{789}{}% Document's page range.
% If second parameter is left empty, the last page is computed automatically.
%\Yearpublication{2006}%
%\Yearsubmission{2005}%
%\Month{11}%
%\Volume{999}%
%\Issue{88}%
% \DOI{This.is/not.aDOI}%

\title{Application of the relativistic precession model to the accreting \\ millisecond X-ray pulsar IGR J17511-3057}

\author{Ivan Zh. Stefanov\thanks{  \email{izhivkov@tu-sofia.bg}\newline}
%Example
%for footnote, note the usage of the \texttt{fnmsep}
%command as separator between institute number and footnote mark}
}
\titlerunning{RP model and IGR J17511-3057}
\authorrunning{I.Zh. Stetfanov}
\institute{Department of Applied Physics, Technical University of Sofia, 8, St. Kliment Ohridski Blvd., 1000 Sofia, Bulgaria
}

%\received{...}
%\accepted{...}
%\publonline{...}


\abstract{%
  The observation of a pair of simultaneous twin kHz QPOs in the power density spectrum of a neutron star or a black hole allows its mass-angular-momentum relation to be constrained. Situations in which the observed simultaneous pairs are more than one allow the different models of the kHz QPOs to be falsified. Discrepancy between the estimates coming from the different pairs would call the used model into question. In the current paper the relativistic precession model is applied to the twin kHz QPOs that appear in the light curves of three groups of observations of the accreting millisecond X-ray pulsar IGR J17511-3057. It was found that the predictions of one of the groups are practically in conflict with the other two. Another interesting result is that the region in which the kHz QPOs have been born is rather broad and extends quite far from the ISCO.}

\maketitle

\section{Introduction}
Quasi-periodic oscillations (QPOs) observed in the power-density spectra of low-mass X-ray binaries (LMXB) have attracted significant research interest due to their potential to serves as a probe of strong field gravity. It is believed that they are intrinsically related to processes that occur in the immediate vicinity of massive objects such as black holes and neutron stars. QPOs are among the few phenomena that could be used for the measurement of the spin and the mass of a neutron star or a black hole. As it was demonstrated in (T\"{o}r\"{o}k et al. 2010; Kotrlov\'{a} et al. 2014) if twin kHz frequencies, a lower and a upper, are observed  simultaneously in the PDS of a LMXB the mass-angular-momentum relation of the central object can be determined. In other words, if the angular momentum of the central object is known from another, independent measurement (not related to QPOs) QPOs allow us to determine its mass. An interesting situation might occur with sources for which more than one pair of twin kHz QPOs is observed. Each of them can be used to determine the mass-angular-momentum relation. Whether the predictions coming from the different pairs are in agreement with each-other depends on the validity of the model that has been used and on the proper interpretation of the observational data. Thus we could test the models proposed the for the explanation of the QPOs. One of the sources for which more that one pair of simultaneous kHz QPOs is observed is the accreting millisecond X-ray pulsar IGR J17511-3057. Kalamkar, Altamirano \& van der Klis (2011) observed three groups of simultaneous twin peak kHz QPOs. In the current paper we apply the relativistic precession (RP) model to this object hoping to learn more about the nature of the central object, which could shed more light on its classification,  and to check the different scenarios proposed by the authors of (Kalamkar et al. 2011) for the identification of kHz QPOs. It turns out that the opposite perspective, according to which the twin kHz QPOs are properly identified and an appropriate model for them is sought, reveals unexpected features of the RP model.

The paper is organized as follows. The observed twin kHz QPOs in IGR J17511-3057 are briefly presented in Section \ref{Observations}.  A brief presentation of the RP model follows in Section \ref{RP}. Section \ref{Ma} is dedicated to the estimates of the mass-angular-momentum ratio coming from the different observations and to the role of uncertainties. The agreement between them is discussed and constraints about the mass of the central object are given. The radii on which the consecutive pairs of kHz QPOs originated are discussed in Section \ref{Section_radii}.  Then come  the Discussion and the Conclusion.

In this paper all the masses are scaled with the Solar mass, the radii are scaled with the gravitational radius $r_{\rm g}\equiv G M/c^2$, the specific angular momentum $a\equiv J/c M^2$ and measure units in which $G=1=c$ are used.
\section{Observations}\label{Observations}

Kalamkar et al. (2011) present the results from 71 pointed observations of IGR J17511-3057 with the RXTE PCA covering the period from September 12 to October 6.
To improve statistics the authors combine close in time and color\footnote{For more details on the color and the classification of the states of X-ray sources we refer the reader to (van der Klis 2006a).} observations into seven groups of 7 to 15. The power spectrum of each group is fitted by a multi-Lorentzian function which is a sum of several Lorentzians. Each Lorentzian corresponds to a different component in the power density spectrum (PDS). For the identification of the different components correlation diagrams (Wijnands \& van der Klis 1999; Psaltis et al. 1999; van Straaten et al. 2005) have been used (See Section 3.3 in (Kalamkar et al. 2011) for more details.). In the current paper only the components recognized as twin kHz QPOs will be considered. Such components have been seen in three of the seven groups of observations - 1, 2 and 7. Group 1 contains 7 observations made between MJD 55087 and MJD 55090, group 2 -- 9 observations made between MJD 55090 and MJD 55092,  and group 7 consists of 14 observations in the period between MJD 55103 and MJD 55111. For each of these groups five Lorentzians were used, i.e. five different components have been recognized in their PDS. The authors of (Kalamkar et al. 2011) found that the averaged light curve of the last five observations of gr. 1 can be fitted also with six Lorentzians. Group 2 can also be alternatively fitted with six instead of five Lorentzians. The alternative fits are designated as 1a and 2a, respectively.

The values of the twin kHz QPOs, their uncertainties, quality factors Q and significances are reproduced here in Table \ref{tab}, where in each group the frequency of the upper kHz QPOs is above that of the lower kHz QPOs. The groups are arranged in order of appearance.

In almost all of the cases $Q\geq2$ which means that these components can be recognized as QPOs. An exception is the lower kHz frequency seen in groups 2 and 2a. The significance of the presented figures is also high.

As Kalamkar et al. (2011) state, there is a number of ways in which the components presented in Table \ref{tab} can be interpreted. The authors propose the following scenarios. \footnote{Here we focus on the kHz QPOs. Details about the other spectral components can be found in (Kalamkar et al. 2011).} Scenarios 1 and 2 propose the same identification for the two high-frequency components. According to these scenarios the high-frequency QPOs seen in groups 1, 2, and 7 are twin kHz QPOs. In scenario 3 the highest frequency QPOs are upper kHz QPOs and the second highest frequency QPOs are hecto Hz QPOs. Scenario 4 identifies the two highest frequency QPOs in groups 1 and 2 as hecto Hz QPOs. In this scenario the two highest frequencies of group 7 cannot be identified.

All of the proposed scenarios have drawbacks. As noted by Kalamkar et al. (2011) the state of the studied object resembles the extreme island state (EIS) observed in accreting millisecond X-ray pulsars and in some atoll sources. In this state neither twin kHz QPOs, nor hecto Hz QPOs  have been seen. Twin kHz QPOs in EIS have been observed only in 4U 1728-34 (Migliari et al. 2003). A significant drawback of scenario 4 is that the absence of upper kHz frequency does not allow the correlations among the frequencies of the power spectral components  to be used for the identifications of the other components.

%\pagebreak%%%%%%%%%%%%%%%%%
\begin{table}
\centering%%%
\caption{HF QPOs observed in the PDS of IGR J17511-3057 }% \citet{Altamirano}}
\label{tab}
\begin{tabular}{cccc}\hline
group & $\nu_{\rm max}$, (Hz)  & Q & significance\\
\hline
\hline
1& $251.8 \pm 13.9$ & $4.3\pm2.8$   & 3.1\\
 & $139.7 \pm 4.2$  & $3.3\pm1.1$   & 3.8 \\
2& $272.2 \pm 13.9$ & $2.45\pm1.6$  & 3.3 \\
 & $129.9 \pm 11.0$ & 1.3           & 3.7\\
7& $179.9 \pm 14.9$ & 2.0           & 4.5\\
 & $72.5 \pm 4.9$&    2.3           & 4.0\\
\hline
1a& $262.1 \pm 18.1$& $2.83\pm2.1$  & 2.7 \\
  & $140.3 \pm 3.4$ & $3.95\pm1.2$  & 4.9 \\
2a& $271.6 \pm 14.5$& 2.0           & 4.4\\
  & $123.0 \pm 8.6$&  $1.2\pm0.4$   & 5.5\\
\hline
\end{tabular}
\end{table}
\section{Relativistic precession model}\label{RP}
The RP model was proposed by Stella \& Vietri (1998; Stella, Vietri and Morsink 1999; Stella and Possenti 2009) as an explanation of the correlation between the low frequency QPOs and the lower kHz QPOs that has been observed by Psaltis, Belloni \& van der Klis (1999; 2002; van der Klis 2006b)  for a large number of neutron-star sources. Later, they applied it also to black holes (Merloni et al. 1999).  This model belongs to the class of the so-called hot spot models according to which the X-ray flux of different compact objects, such as black holes and neutron stars, is  modulated by the motion of matter inhomogeneities in the inner region of the accretion disk where most of the energy is released. The RP model treats the hot spot as a free test particle and attributes the frequencies that appear in the power density spectra to its geodetic orbital and epicyclic frequencies. The horizontal branch oscillations $\nu_{\rm h}$, the lower $\nu_{\rm l}$ and the upper $\nu_{\rm u}$ twin kHz QPOs are ascribed to the nodal precession $\nu_{\rm nod}=|\nu_{\rm \phi}-\nu_{\rm \theta}|$, the periastron precession $\nu_{\rm per}=\nu_{\rm \phi}-\nu_{\rm r}$ and the orbital $\nu_{\rm \phi}$ frequencies, respectively. Here $\nu_{\rm r}$ and $\nu_{\rm \theta}$ are the radial and the vertical epicyclic frequencies. They have been obtained for the first time for a black hole in (Aliev \& Gal'tsov 1981; Aliev, Gal'tsov \& Petukhov 1986). Explicit formulas for them in the special case of Kerr space-time can be found in the Appendix \ref{appendix}.

T\"{o}r\"{o}k et al. (2010) apply the RP model to estimate the mass of the central neutron star in Circinus X-1. The authors show that such an estimate results in a specific mass-angular-momentum relation. Recently, following the same approach Kotrlov\'{a} et al. (2014)  studied the atoll source 4U~1636-53. The RP model has been also recently applied in (Motta et al. 2014a; Motta et al. 2014b; Bambi 2013) for the precise measurement of the mass and angular momentum of black holes.
\section{Mass-angular-momentum relations}\label{Ma}
\subsection{The $a-M$ diagram}
If a pair of simultaneous kHz QPOs is observed and a model for them is chosen then the following two equations can be composed
\begin{eqnarray}
&&\nu_{\rm l}(a,M,r)=\nu_{\rm l}^{\rm obs} \label{basic1}\\
&&\nu_{\rm u}(a,M,r)=\nu_{\rm u}^{\rm obs}.\label{basic2}
\end{eqnarray}
The system (\ref{basic1})-(\ref{basic2}) has two equations for three variables. If we fix one of the variables we can solve it for the other two.
Another way of saying this is that the system (\ref{basic1})-(\ref{basic2}) defines two of the variables as functions of the third variable in implicit
form. If we fix the value of the radius of the orbit along which the hot spot propagates we can find the values of the mass and the angular momentum of
the central object. If $r$ is varied another pair of values of $a$ and  $M$ is obtained. In other words we obtain a 1-parameter family of solutions. The two functions $a(r)$ and $M(r)$ define a curve in the $a-M$ space in parametric form.
\subsection{Uncertainties}
When the uncertainties of $\nu_{\rm l}^{\rm obs}$ and $\nu_{\rm u}^{\rm obs}$ are taken into account the curve described in the preceding subsection becomes a stripe. (Different values of these frequencies would yield different estimates for $a$ and $M$.) In other to build the stripe we need the uncertainties $\Delta a$ and $\Delta M$. They can be found from the following uncertainty propagation formulas
\begin{eqnarray}
\Delta\nu_{\rm u}^{\rm obs}=\left|{\partial \nu_{\rm u} \over \partial a}(a(r),M(r),r)\right|\Delta a(r)\\ \nonumber +\left|{\partial \nu_{\rm u} \over \partial M}(a(r),M(r),r)\right|\Delta M(r)\\
\Delta\nu_{\rm l}^{\rm obs}=\left|{\partial \nu_{\rm l} \over \partial a}(a(r),M(r),r)\right|\Delta a(r)\\ \nonumber +\left|{\partial \nu_{\rm l} \over \partial M}(a(r),M(r),r)\right|\Delta M(r).
\end{eqnarray}
The uncertainties $\Delta\nu_{\rm l}^{\rm obs}$ and $\Delta\nu_{\rm u}^{\rm obs}$ are taken for each frequency from Table \ref{tab}.
The coefficients in the above system of linear equations depend on $r$, so, again, for $\Delta a$ and $\Delta M$ a 1-parameter family of solutions is obtained.

The stripes corresponding to the five pairs of kHz QPOs given in Table \ref{tab} are presented and discussed  bellow.
\subsection{On the equivalence of groups 1 and 1a, and groups 2 and 2a}
\begin{figure}
\center
\includegraphics[width=0.4\textwidth]{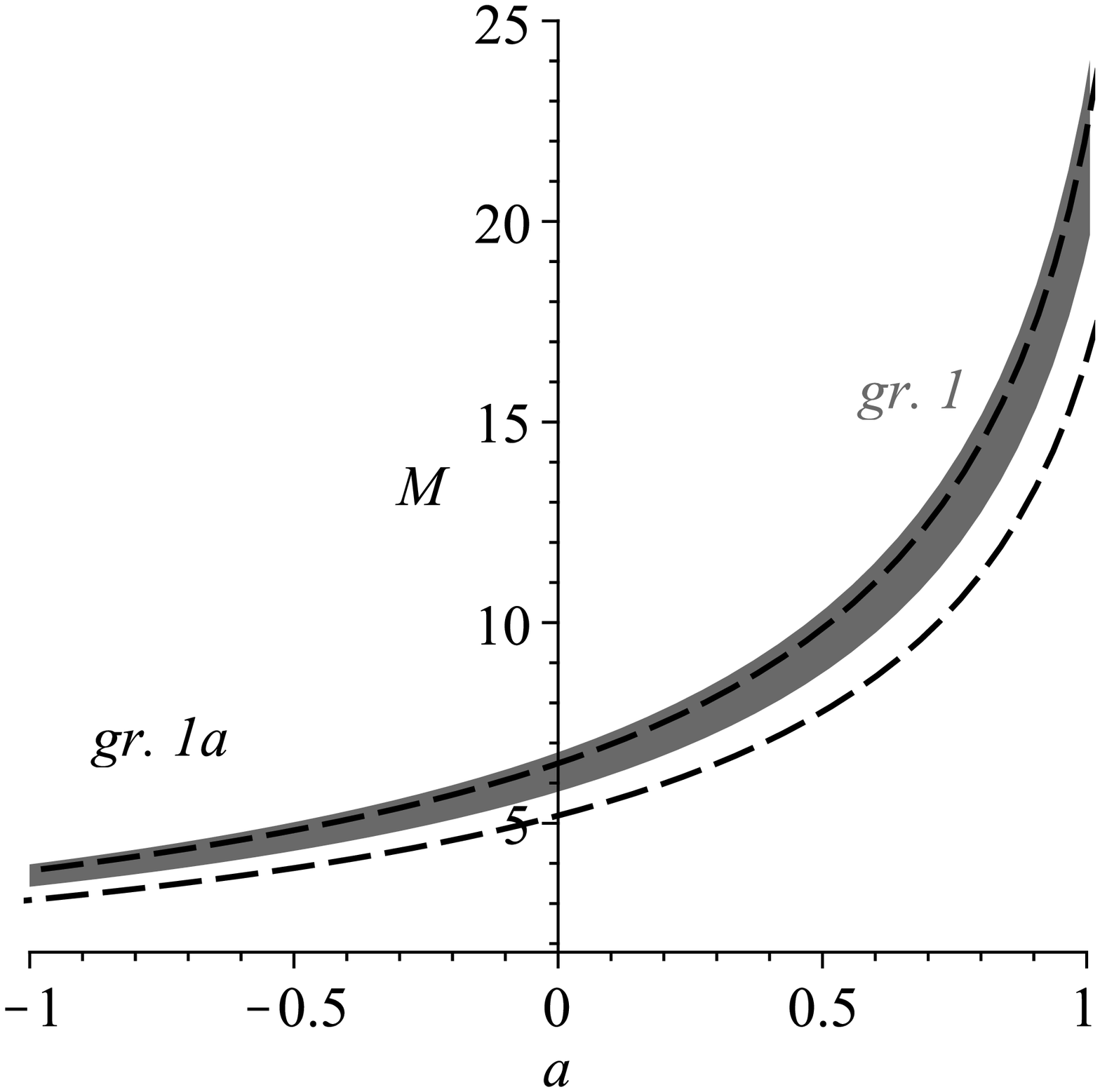}
\includegraphics[width=0.4\textwidth]{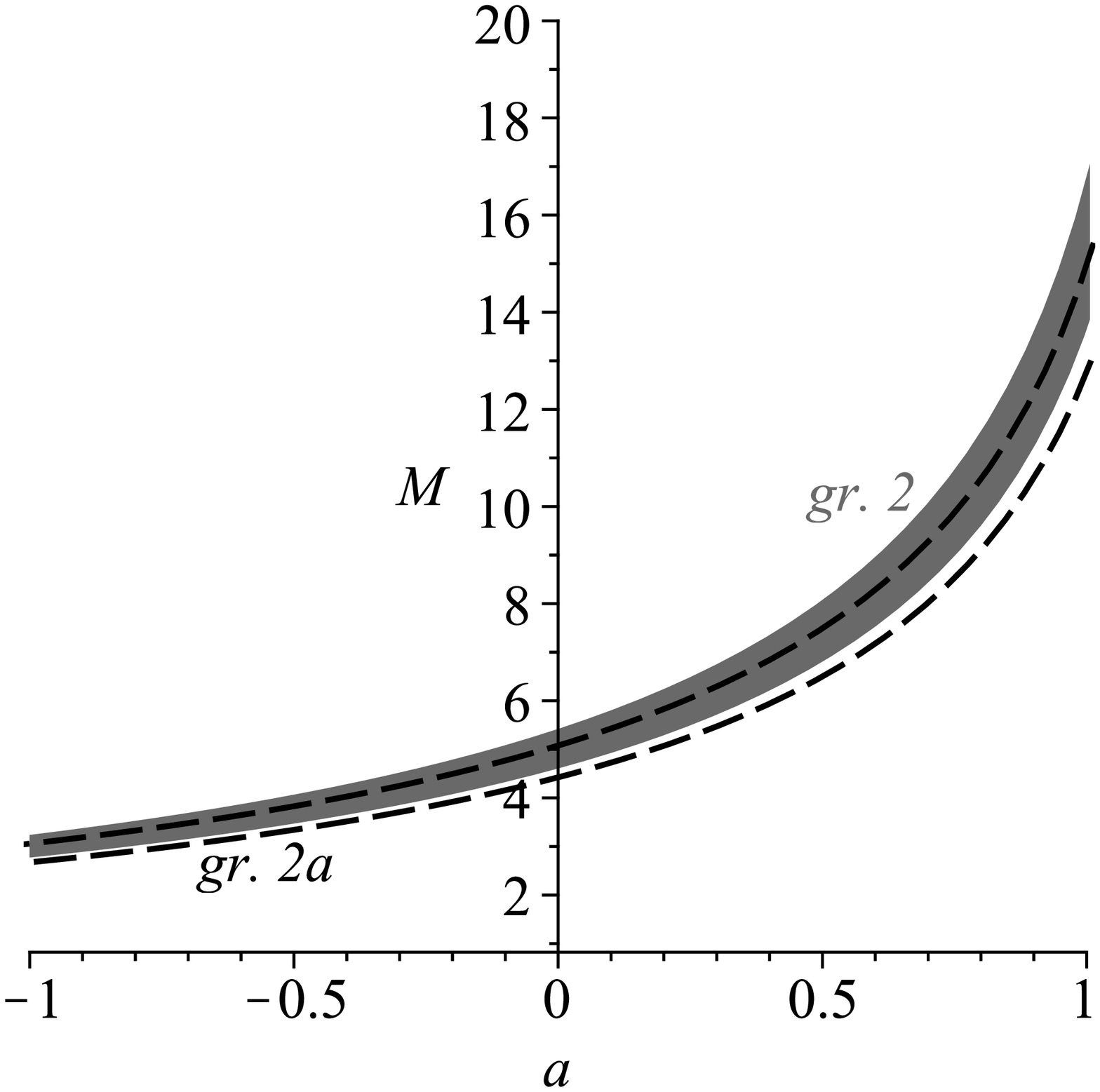}
\caption{\textit{top}: The predictions for the $a-M$ relation coming from groups 1 and 1a. The former is represented by a filled gray stripe, while the latter is given by the empty stripe enclosed between the two dashed lines. \textit{bottom}: The same as in the top panel but for groups 2 and 2a.}
\label{equivalence}
\end{figure}

The six-Lorentzian fits 1a and 2a have different content in the low-frequency complex than 1 and 2. The high-frequency peaks of the former are also shifted with respect to the latter. One would naturally ask whether the alternative fits, 1 and 1a, and, 2  and 2a, would yield the same mass-angular-momentum relations. The corresponding regions in the $a-M$ parameter space are shown on Figure \ref{equivalence}. The constraints of groups  1 and 1a are presented on the top panel of the figure, while those of groups 2 and 2a -- on the bottom panel. The estimates coming from gr. 1 are represented by a filled dim gray stripe. The stripe of gr. 1a is empty and is enclosed by two dashed lines. The same designation is chosen for the bottom panel, i.e. for groups 2 and 2a. As it can be seen, the two stripes -- the filled one and the empty one, partially overlap for the hole range of values of the angular momentum. In other words, the constraints on $a$ and $M$ coming from the alternative fits are in agreement with each-other. Hence, below we choose to work only with groups 1 and 2.
\subsection{On the agreement of the predictions of the three pairs -- groups 1, 2 and 7}
\begin{figure}
\center
\includegraphics[width=0.4\textwidth]{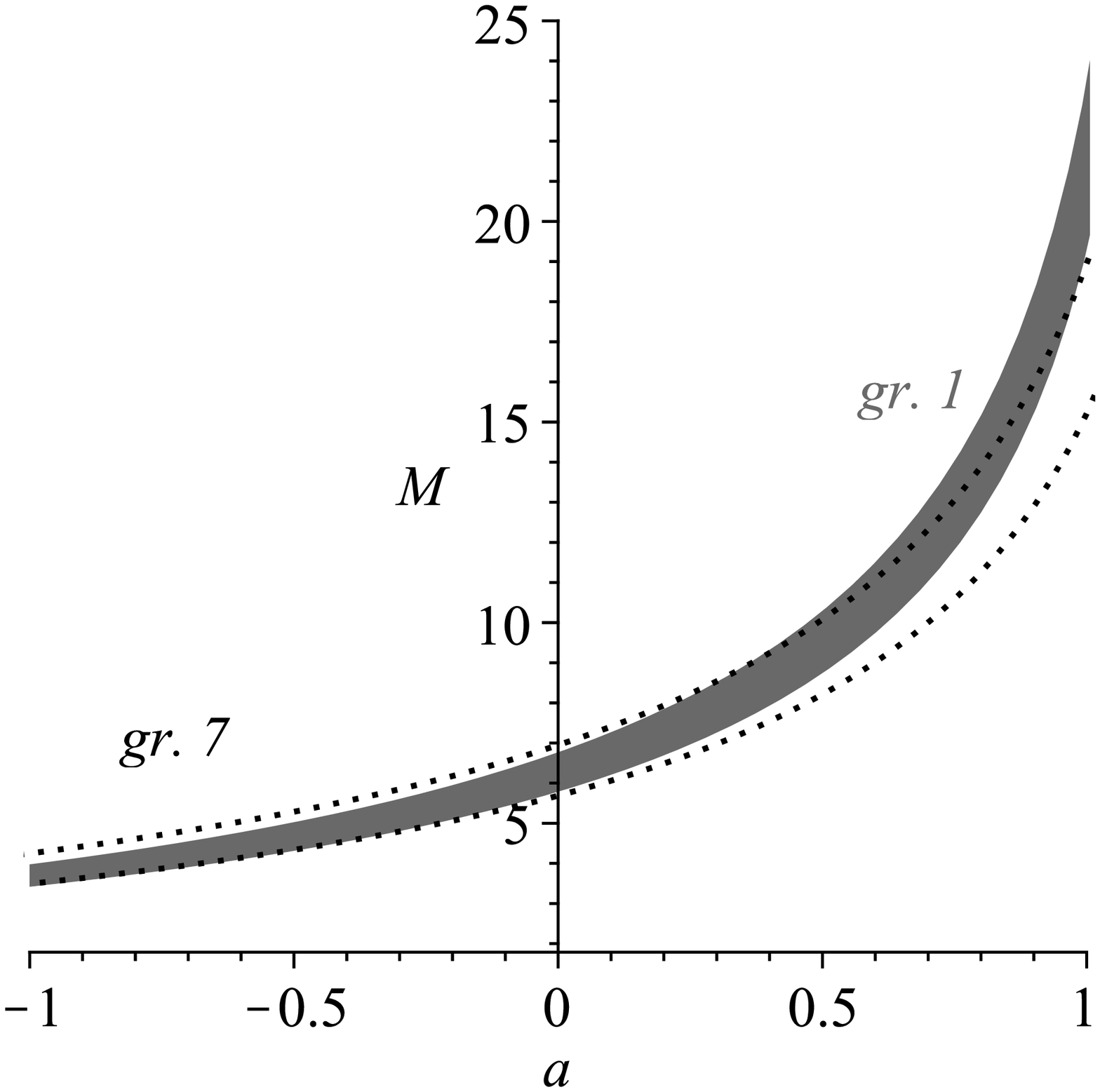}
\includegraphics[width=0.4\textwidth]{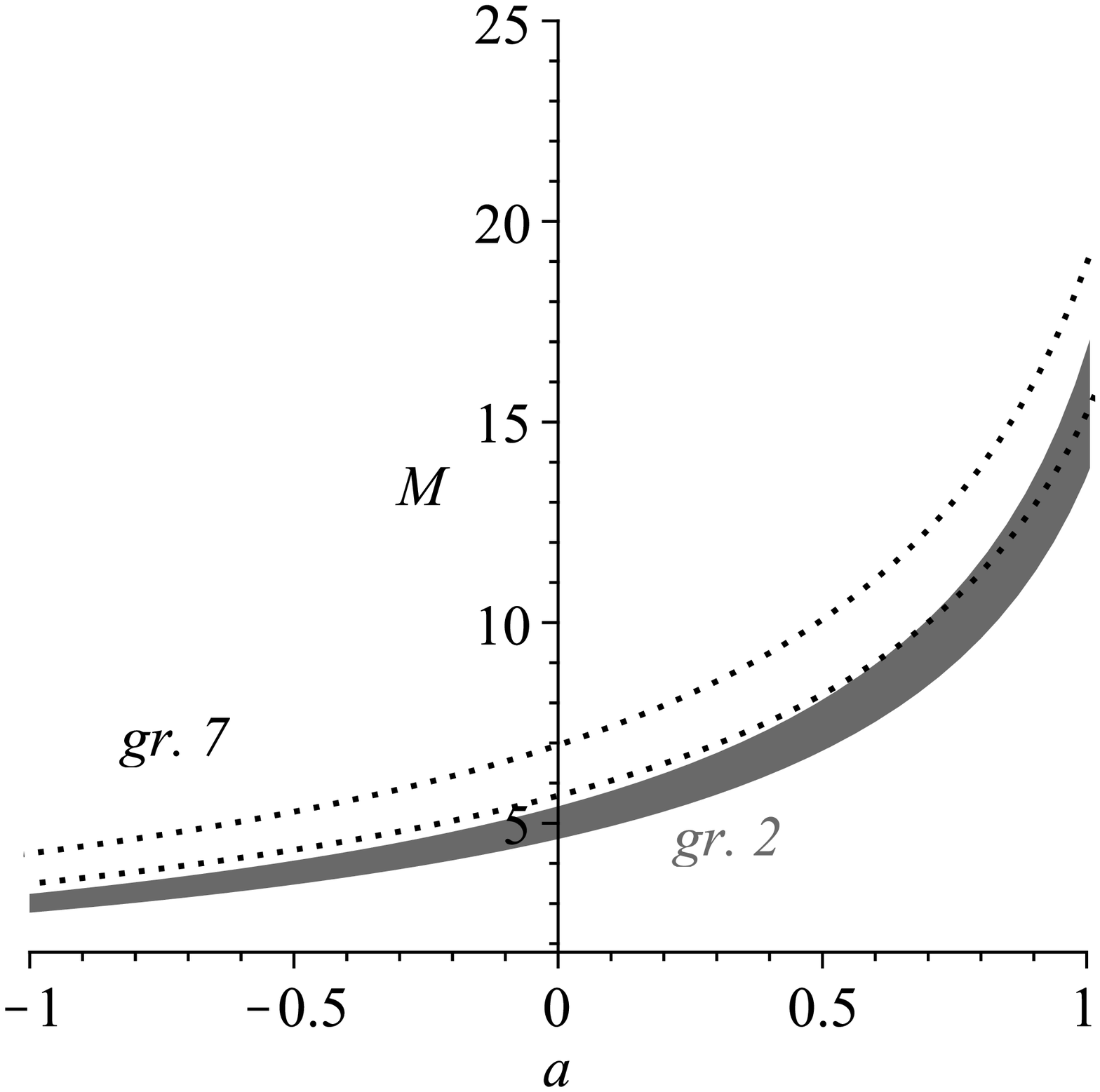}
\includegraphics[width=0.4\textwidth]{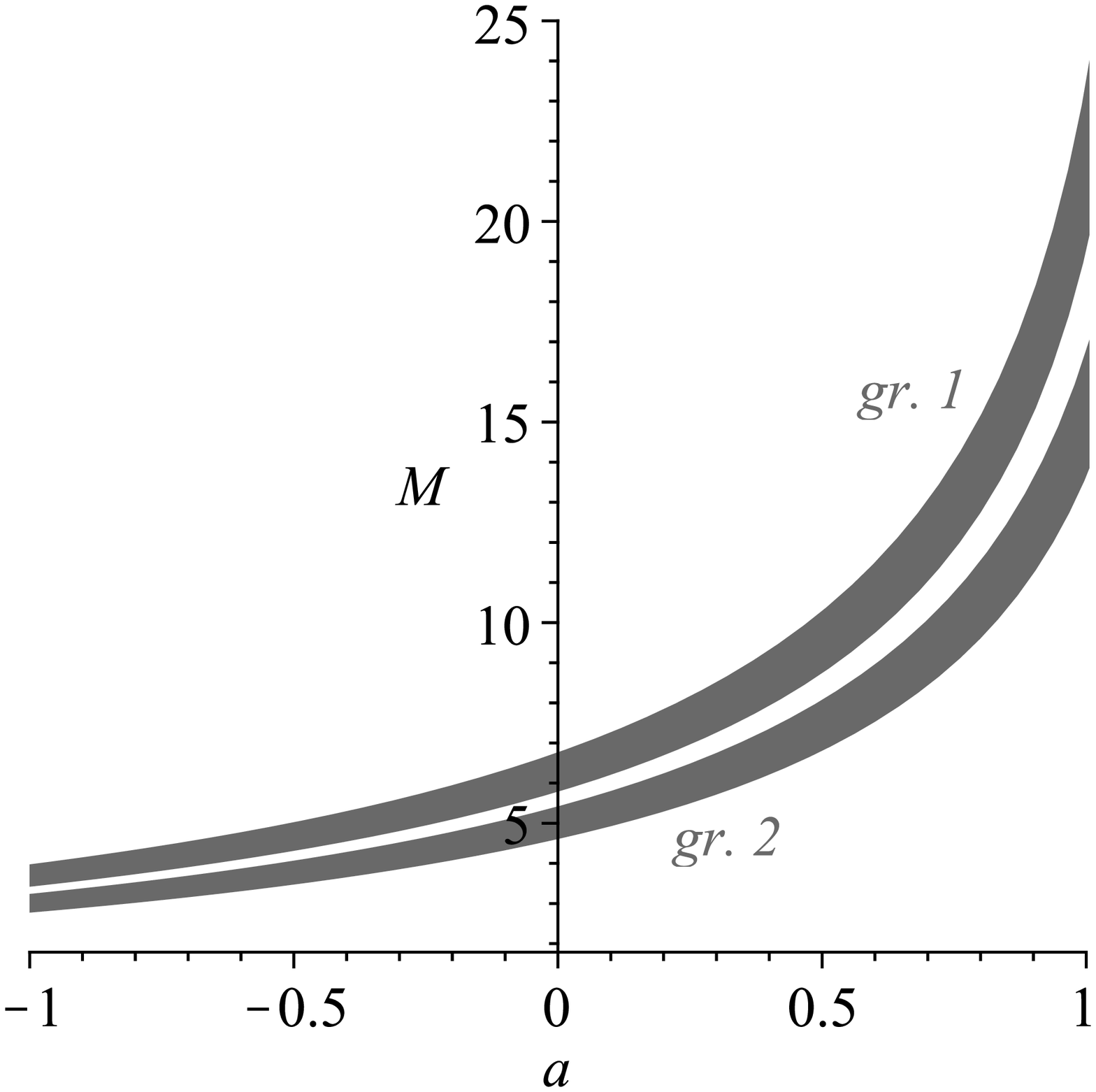}
\caption{\textit{top}: The predictions for the $a-M$ relation coming from groups 1 and 7. The former is represented by a filled gray stripe, while the latter is given by the empty stripe enclosed between the two dotted lines.  \textit{middle}: The same as in the top panel but for groups 2 and 7. \textit{bottom}: The same as in the top panel but for groups 1 and 2.}
\label{ageement}
\end{figure}

As a next step we check whether the $a-M$ relations coming from the three groups in which simultaneous twin peaks of kHz QPOs are observed, namely groups 1, 2 and 7,  are in agreement with each other. They are compared on Figure \ref{ageement}. Here and bellow we allow $a$ to vary in the interval $[-1,1]$ since bigger absolute values are less likely. On Figure \ref{ageement} groups 1 and 2 are represented by filled dim gray stripes while the stripe of group 7 is empty and enclosed by two dotted lines. The estimates of groups 1 and 7 are compared on the top panel, 2 and 7 on the middle panel, and 1 and 2 on the bottom panel. It appears that groups 1 and 7 give conforming results for almost all values of the angular momentum with the exception of a narrow interval of  high values of $a$. The stripes of groups 2 and 7 overlap only for a small interval of $a$. The stripe of groups 1 and 2 are completely detached. Some possible explanations are given below in the Discussion.
\subsection{Estimates for $M$ }
Higher values of $r$ yield lower masses and angular momenta\footnote{According to the RP model the $M$ and $a$ are positively correlated. This fact has been noted also in (T\"{o}r\"{o}k et al. 2010) and, as the authors state, it is challenging for the RP model in the case of prograde rotation. We should clarify, however, that in the cited paper a different convention for the angular momentum has been chosen -- it is a positive quantity and the cases of prograde and retrograde rotation of the matter in the accretion disk are treated separately (with different formulas for each case). In the current paper we allow $a$ to take both positive and negative values corresponding to prograde and retrograde rotation of the accretion disk, respectively, so here $a$ and $M$ are positively correlated, while $|a|$ and $M$ are positively correlated only for the prograde orbits.   }. Since $a$ cannot be lower that $-1$ and higher than $1$ we can find a lower and a upper bound on the mass of the object. The estimates coming from the three groups are  $3.7\leq M_{\rm gr. 1}/M_\odot\leq21.3$, $3.0\leq M_{\rm gr. 2}/M_\odot\leq15.1$ and $3.9\leq M_{\rm gr. 7}/M_\odot\leq17.0$.
\section{Radii on which the observed twin kHz QPOs were born}\label{Section_radii}
\subsection{Radii}
\begin{figure}
\center
\includegraphics[width=0.4\textwidth]{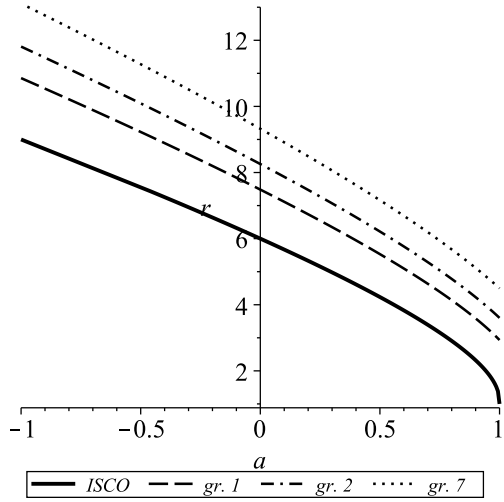}
\caption{The radii on which the kHz QPOs of groups 1, 2 and 7 were born.}
\label{radii}
\end{figure}

The $a(r)$ and $M(r)$ relations obtained from system (\ref{basic1})-(\ref{basic2}) can be considered also in the following way. The $a(r)$ function can be inverted. The inverse functions tell us on what radius the kHz QPOs from a given group were born for given $a$. The results are presented on Figure \ref{radii}. The first thing that can be seen that for all the three groups, 1, 2 and 7, the kHz QPOs were born outside of the ISCO, as we would expect. A less expected result is that the groups which were observed earlier in time are associated with inner radii than the later groups. Actually for the whole considered interval of values of the angular momentum $r_{\rm gr. 1}<r_{\rm gr. 2}<r_{\rm gr. 7}$.

\begin{figure}
\center
\includegraphics[width=0.4\textwidth]{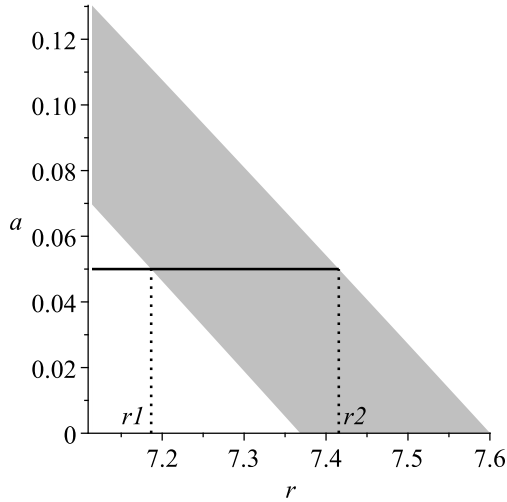}
\caption{The $a(r)\pm\Delta a(r)$ stripe.}
\label{Dr}
\end{figure}

\begin{figure}
\center
\includegraphics[width=0.4\textwidth]{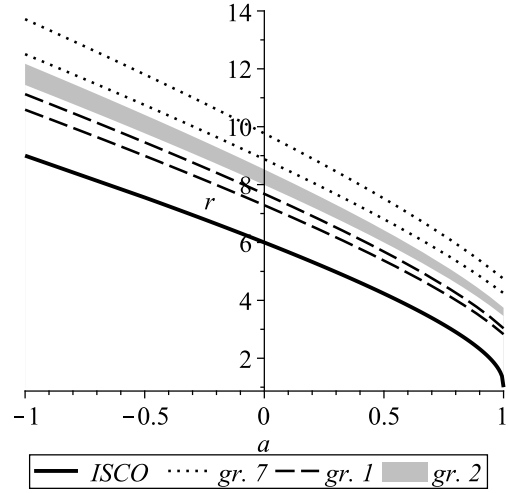}
\caption{The radii on which the kHz QPOs of groups 1, 2 and 7 were born with error bars. The two outer stripes are blank and are presented only by their boundaries.}
\label{radii_stripes}
\end{figure}

A natural question that comes next is whether this result will be confirmed if the uncertainties of the radii are taken into account. In other to evaluate $\Delta r$ we can again use the $a(r)$ relation. Due to the uncertainty $\Delta a(r)$, to each value of the radius $r$ corresponds an interval of values of $a$. The inverse is also true -- each value of $a$ yields an interval of values of $r$. This is shown on Figure \ref{Dr}. In this example $a=0.05$, $r_1=7.187$ and $r_2=7.416$. Putting $r_1=r-\Delta r$ and $r_2=r+\Delta r$ we obtain
\begin{equation}
\Delta r (a)={r_2(a)-r_1(a)\over 2}.
\end{equation}
The stripes $r(a)\pm\Delta r(a)$ are given on Figure \ref{radii_stripes}. As it can be seen, even when the uncertainties of $r$ are taken into account it occurs that the earlier groups are related to inner regions of motion of the hot spots.
%In other words it appears that, contrary to what one would expect,  the hot spot has moved outwards.

\subsection{Energy of the hot spot}
The hot spots which modulate the X-ray flux cover a broad region of the accretion disk. Hot spots orbiting at larger radii have greater energy. As a demonstration of the significance of the width of the region involved in the modulation of the emission we can evaluate the energy difference between the inner and the outer hot spots.

The metric of a general stationary spacetime takes the following form in Boyer-Lindquist
\begin{eqnarray}
ds^2 =   g_{tt} dt^2 + g_{rr}dr^2 + g_{\theta\theta} d\theta^2\\
+ 2g_{t\phi}dt d\phi + g_{\phi\phi}d\phi^2.
\end{eqnarray}
The energy per unit mass $E$ of a test body propagating along a circular geodesic in such spacetime is given by the standard formula
\begin{equation}
E = - \frac{g_{tt} + g_{t\phi}\Omega}{
\sqrt{-g_{tt} - 2g_{t\phi}\Omega - g_{\phi\phi}\Omega^2}},
\end{equation}
where $\Omega$ is the angular velocity of the body, defined as
\begin{eqnarray}
\Omega&=& \frac{d\phi}{dt} = \\ \nonumber
&&\frac{- \partial_r g_{t\phi}
\pm \sqrt{\left(\partial_r g_{t\phi}\right)^2
- \left(\partial_r g_{tt}\right) \left(\partial_r
g_{\phi\phi}\right)}}{\partial_r g_{\phi\phi}}
\end{eqnarray}
For the particular case of Kerr space-time the energy is
\begin{equation}
E ={\frac {{r}^{2}-2\,r+a\sqrt {r}}{r\sqrt {{r}^{2}-3\,r+2\,a\sqrt {r}}}}.
\end{equation}

\begin{figure}
\center
\includegraphics[width=0.4\textwidth]{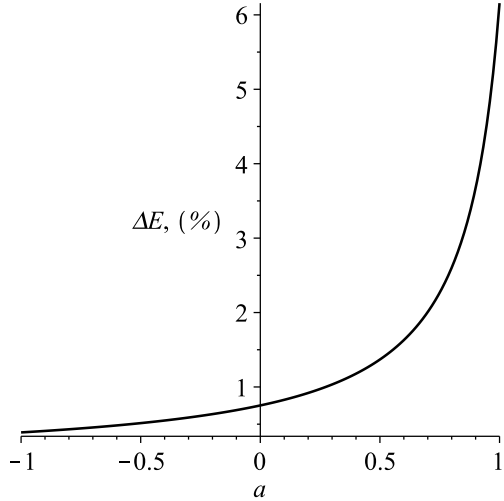}
\caption{The increase of the energy $\Delta E$ of the hot spot during its evolution between $r_{\rm gr. 1}$ and $r_{\rm gr. 7}$  expressed in percents as function of the angular momentum $a$ of the central object.}
\label{Energy}
\end{figure}

The energy difference between $r_{\rm gr. 1}$ and $r_{\rm gr. 7}$ in percents can be calculated by the formula
\begin{equation}
\Delta E(a) ={E\left(a,r_{\rm gr. 7}(a)\right)-E\left(a,r_{\rm gr. 1}(a)\right)\over E\left(a,r_{\rm gr. 1}(a)\right)}\times 100.
\end{equation}
The dependence of $\Delta E$ on $a$ is shown on Figure \ref{Energy}. For $a$ varying in the interval $[-1,1]$, $\Delta E\in[0.4,6.2]$. The significance of these figures is discussed bellow.

\section{Discussion}
There are several issues that we would like to comment on: the conflicts between the predictions for the $a-M$ relation of gr. 1 and gr. 2 and of gr. 7 and gr. 2, the high mass estimates and the fact that the twin kHz QPOs that occurred later were born on outer radii. There are three factors which determine the estimates for the mass-angular-momentum relation of the studied object IGR J17511-3057 -- the choice of a metric (a model for the spacetime in the vicinity of the neutron star), the identification of the twin QPOs and the choice of a model for them. They are interrelated and tested simultaneously. The reason for the issues ascertained above, might be a failure in any of these factors. The current stage of the study does not allow us to determine what is their individual contribution.

Now let us comment on the choice of the metric. In general, analytical metrics are more convenient for astrophysical applications than
numerical ones. An example is the Hartle-Thorne metric which has three free parameters -- mass $M$, specific angular momentum $a$ and dimensionless
quadrupole momentum $q$. It can be applied for the description of the external geometry of slowly rotating neutron stars whose rotation frequency is
smaller than the mass-shedding frequency ($f_{\rm mass \,\,shedding} \sim 1100$ Hz) and dimensionless spin $|a|< 0.4$ (Hartle 1967; Hartle \& Thorne 1968; Chandrasekhar \& Miller 1974; Miller 1977; Urbanec, Miller \& Stuchl\'{\i}k 2013; T\"{o}r\"{o}k et al. 2010; Stuchl\'{\i}k \& Kolo\v{s} 2015 and references therein). The constraint for the rotation frequency is satisfied by IGR J17511-3057 since the spin frequency cited by Kalamkar et al. (2011) is 244.8 Hz. The deviations between the Hartle-Thorne metric and more realistic numerical metrics such as those obtained by the LORENE or RNS codes is within few percent for near-maximum-mass neutron stars with $|a|<0.4$.

When $|a|<0.4$ and $q/a^2<2$ the Hartle-Thorne geometry coincides with the Kerr geometry at the level
of about few percent. As it is shown in (Urbanec, Miller \& Stuchl\'{\i}k 2013) such low values of $q/a^2$ are observed for near-maximum-mass neutron stars for equations of state allowing maximum masses higher than $2M_{\odot}$. The Kerr metric is rather attractive due to the simplicity of the analytic expressions  and the fact that it has only two free parameters, $a$ and $M$. This is why we choose to work with it but we should note that this choice is justified only for slowly rotating neutron star neutron stars with dimensionless spin $|a|<0.4$ and mass close to or higher than $2M_{\odot}$. Here we have worked with a wider range of values of $a$, $a\in[-1,1]$, for completeness. The application of the Kerr metric is also legitimate for objects with $M>3$ but since this is the theoretical upper bound on the neutron star mass so heavy objects cannot
be regarded as neutron stars.
%It can be applied for the estimation of the masses of black holes but for the case of neutron stars, however, this is questionable.

The choice of the Kerr metric in the current study in practice means that we make the hypothesis that the central object is a
near-maximum-mass slowly rotating neutron star. This hypothesis is supported by the observation that the values of the twin QPOs given in Table \ref{tab} are relatively low. Actually their magnitude is close to the twin HF QPOs observed in the black--hole binaries GRS 1915+105 and XTE 1550-564 (McClintock \& Remillard 2006). Taking this into account and the fact that according to the RPM the QPO frequencies are inversely proportional to the masses of the objects one might be tempted to expect that the mass of IGR J17511-3057 is close to the masses of these black holes. We should note, however, that the magnitude of the twin QPOs alone does not provide sufficient information for a lower bound on the mass to be found -- the radii are also needed. Low values of the frequencies could be explained not only by big masses but also by big radii. The ratio or, which is equivalent, the difference of the frequencies also plays significant role.

The question that comes next is wether the results that we obtain are in conflict with our hypothesis. Masses significantly lower than $2M_{\odot}$ would be in conflict with it. Masses greater than $3M_{\odot}$ are also problematic. They are too high for a neutron star and give us a reason to question not only the metric but also the model for the twin kHz QPOs and interpretation of the observational data. For masses in the
interval $2-3M_{\odot}$ we have no obvious reason to question any of these points.
%The discrepancies between this metric and realistic neutron star metrics become less significant for high masses ((T\"{o}r\"{o}k et al. 2010; Stuchl\'{\i}k \& Kolo\v{s} 2015 and references therein).

%\textbf{In the absence of prior knowledge of the mass and spin of the studied object we cannot be sure that the proper metric has been chosen. There are some indications, however, that the use of the Kerr metric for the description of the spacetime in the vicinity of IGR J17511-3057 is not so groundless.}

The lowest mass estimate given here comes from group 2. It is $3M_{\odot}$. This mass justifies the use of the Kerr metric but is unnatural for neutron stars. One would like to know how robust is this result to the values of the frequencies of the twin kHz QPOs and the angular momentum. If we assume that the RP model provides an adequate explanation of the observed QPOs and that the latter have been correctly identified with the frequencies given in Table \ref{tab}, $M < 3$ requires $a<-1$ and $M < 2$ yields $a<-2$. Such angular momenta are unrealistic.

%\textbf{One would like to know what is the role of $a$ for the mass estimation. The application of the Kerr metric would be questionable if the mass of the object was less than $ 2M_\odot $.}

Now, let us check what values of the frequencies would yield $3M_\odot $ for $-0.4\leq a\leq 0.4$, the range in which the Kerr metric is legitimately used? We consider the following system
\begin{eqnarray}
\nu_{\rm l}(a,M=3,r)&=&x,\label{spec1}\\
\nu_{\rm u}(a,M=3,r)&=&{\nu_{\rm u}^{\rm obs}\over \nu_{\rm l}^{\rm obs}}\,x \label{spec2}.
\end{eqnarray}
We require that the new frequencies have the same ratio as those cited in Table \ref{tab}.
Varying $a$ in the interval $[-0.4,0.4]$ and changing the groups of observations we obtain different values of $r$ and $x$. It appears that $M= 3 $ requires values of $\nu_{\rm l }$ which are at least $30-70$ \% higher than the observed ones while the uncertainty of this frequency is within only 1\%.

It appears that the problem with the big masses cannot be resolved by a small shift in $\nu_{\rm l }$ (and $\nu_{\rm u }$) or $a$. We should note also that the discrepancy between the theoretical bound for the mass of a neutron star, about $3M_\odot$, and the values of the masses obtained here is  not so severe for moderate absolute values of the angular momentum $|a| < 0.4$. In this case they differ by a factor of few. The question whether more adequate masses would be obtained with a more realistic model for the spacetime calls for further study.
Some brief comments on the other two factors, the interpretation of the observed kHz QPOs and the model chosen for their description, follow.

If we assume that the Kerr metric and the RP model are adequate for the description of the QPOs phenomenology of IGR J17511-3057 then the discrepancy between the expected and the obtained masses and the conflict between the constraints on the $a-M$ relation coming from the different groups of observations could simply mean that one or both of the frequencies in the pairs cannot be identified as high frequency QPOs. In other words, it is possible that neither scenario 1 nor scenario 2 from (Kalamkar et al. 2011) is realized and one of the other two scenarios should be considered.

The applicability of the RP model to this source might also be questioned. A good alternative might be the resonant switch mechanism proposed by Stuchl\'{\i}k, Kotrlov\'{a} \& T\"{o}r\"{o}k (2012). Indeed, as it can be easily seen,  $\nu_{\rm u}/\nu_{\rm l}<2$ for gr. 1 and $\nu_{\rm u}/\nu_{\rm l}>2$ for gr. 2 and gr. 7 which might be an indication that at the resonant point, $\nu_{\rm u}/\nu_{\rm l}=2$, somewhere between gr. 1 and gr. 2,  a switch between different oscillation modes has occurred. In other words, one pair of modes must be applied to gr. 1 and another pair to groups 2 and 7. This scenario, however, needs an answer to the question why the predictions of gr. 1 and gr. 7 overlap so well since the same mode has been used for both of them. In a future work the author plans to consider other models for the HF QPO, e.g. the non-linear resonance model and the resonant switch model.

The fact that $\nu_{\rm u}/\nu_{\rm l}\sim2$ is an indication that the kHz QPOs have originated at an orbit which is relatively far from the ISCO as it was noted in  (T\"{o}r\"{o}k et al. 2010).

We should comment also on the results for the radii on which the kHz QPOs were born. As we have already mentioned, it appears that the groups that occurred earlier in time correspond to inner orbits. Even though gr. 2 gives mass and angular momentum estimates that are in conflict with those of groups 1 and 7, the radii coming from it also fit well in this picture since $r_{\rm gr. 1}<r_{\rm gr. 2}<r_{\rm gr. 7}$ for all the considered values of $a$. These results would be really unexpected if we assume that the QPOs in the three groups of observations are produced by the same hot spot. It seems to have moved outwards during its evolution. To quantify the significance of the outward motion of the hot spot we have evaluated the increase of its  energy in the transition between $r_{\rm gr. 1}$ and $r_{\rm gr. 7}$ (see Figure \ref{Energy}). It is approximately $10-20$ \% of the energy difference between the ISCO and infinity, which is much. According to (Schnittman 2005), however, the lifetime of the hot spots which modulate the X-ray flux is confined to no more than several orbital periods (less than a second) which means that the three pairs of twin kHz QPOs could not be produced by the same hot spot. Even if we give up the single hot spot hypothesis the results are still very interesting since they demonstrate that the origin of the kHz QPOs is related to a rather broad region in the accretion disk and which for the given case extends quite far from the ISCO.

\section{Conclusion}
In the current paper the relativistic precession model is applied to the twin kHz QPOs that appear in the light curves of three groups of observations of the accreting millisecond X-ray pulsar IGR J17511-3057. When applied to a pair of frequencies the RP model allows constraints on the mass-angular-momentum relation of the central object to be posed. It was found that the estimates coming from the two alternative fits of the light curves of the first two groups, namely  1 and 1a, and 2 and 2a, with five and six Lorentzian functions, respectively, are in agreement. The regions which represent the predictions of groups 1 and 7 in the $a-M$ space overlap for almost all of the considered values of the angular momentum with the exception of a narrow interval of very high values of $a$. The predictions of group 2 are practically in conflict with those of groups 1 and 7.
The mass of the central object was found to be greater than $3M_\odot$ which is a too high value for a typical neutron star. The possible reasons for these conflicts, the discrepancy between the predictions of the different groups and the untypically high masses, are: an improper choice of a metric, an improper choice of a model for the kHz QPOs or an incorrect interpretation of the observed frequencies. Had the mass estimates been lower than $2M_{\odot}$ we would have had good reason to suggest that the Kerr metric is not appropriate. This however is not the case and we are inclined to believe that the other two factors are responsible for the issues.  Either the RP model is not applicable to the given object, or one or both of the observed frequencies cannot be classified as kHz QPOs. The latter means that the current study disfavors scenarios 1 and 2 given in (Kalamkar et al. 2011).

One might propose that group 2, or more precisely the point in the $\nu_{\rm l }-\nu_{\rm u }$ plane that corresponds to it, is an outlier. If this group is not taken into consideration that would solve the problem with the conflict of the predictions for the $a-M$ relation but not the problem with the untypically high masses.

The current study also reveals that the QPOs which were observed earlier in time were born, according to the RP model, at inner radii than the later ones. The hot spots which produced the later kHz QPOs have had significantly greater energy than those responsible for the earlier ones.

The study of the horizontal branch oscillations in the context of the relativistic precession model is left for future work.

\acknowledgements
This work was partially supported by the Bulgarian National Science Fund under Grant No DMU 03/6. The author would like to thank Dr. Sava Donkov for the numerous fruitful discussions and the anonymous reviewer for the valuable comments which significantly improved the manuscript.

\newpage%%%%%%%%%%%%%%%%%%%%%%%%%%%%%%%%%%%%%%%%%%%%%%%%%%%%%%

\appendix
\section{Fundamental frequencies}\label{appendix}
The explicit form of the orbital frequency $\nu_{\rm \phi}$ and the two epicyclic frequencies -- the radial $\nu_r$ and the vertical $\nu_{\theta}$ -- for the Kerr black hole can be found, for example, in (Aliev, Esmer \& Talazan 2013)
\begin{equation}
\nu_{\rm \phi} =\left({1\over 2\pi}\right)\frac{ M^{1/2}}{ r^{3/2} + a M^{1/2}}\,\,,
\label{orbf}
\end{equation}
\begin{equation}
\nu_{r}^2 = \nu_{\rm \phi}^2\, \left( 1-\frac{6 M}{r} -\frac{3
a^2}{r^2} + \, 8 a {M^{1/2}\over r^{3/2}}
\right),
\label{kerrradf}
\end{equation}
\begin{equation}
\nu_{\theta}^2= \nu_{\rm \phi}^2\, \left(1
+\frac{3 a^2}{r^2} - \, 4 a {M^{1/2}\over r^{3/2}} \right).
\label{kerraxif}
\end{equation}
A change in the orientation of the orbit (direction of rotation of the hot spot) is equivalent to a change of the direction of rotation of the central object, i.e. a change in the sign of $a$.
\end{document}